\documentclass[10pt]{IEEEtran}

\usepackage{subfiles}
\usepackage{amsthm,amsmath,amssymb}
\usepackage{algorithm,algpseudocode,algorithmicx}
\usepackage{graphicx, subfigure}
\usepackage{cite}
\usepackage{float}

\begin{document}
\title{Unsupervised Change Detection in Satellite Images\\ with Generative Adversarial Network}

\author{Caijun~Ren, Xiangyu~Wang, Jian~Gao,
        and~Huanhuan~Chen,~\IEEEmembership{Senior Member,~IEEE}
\thanks{Caijun Ren, Xiangyu Wang and Huanhuan Chen are with UBRI, School of Computer Science and Technology, University of Science and Technology of China (USTC), Hefei, 230027, China, email: raggie@mail.ustc.edu.cn, sa321@mail.ustc.edu.cn, hchen@ustc.edu.cn. Jian Gao is with StarGIS Technology Development Co.,Ltd, Tianjin, China.}}

%



\maketitle
\IEEEpeerreviewmaketitle

\begin{abstract}
Detecting changed regions in paired satellite images plays a key role in many remote sensing applications.  
The evolution of recent techniques could provide satellite images with very high spatial resolution (VHR) but made it challenging to apply image coregistration, and many change detection methods are dependent on its accuracy.
Two images of the same scene taken at different time or from different angle would introduce unregistered objects and the existence of both unregistered areas and actual changed areas would lower the performance of many change detection algorithms in unsupervised condition.
To alleviate the effect of unregistered objects in the paired images, we propose a novel change detection framework utilizing a special neural network architecture --- Generative Adversarial Network (GAN) to generate many better coregistered images. 
In this paper, we show that GAN model can be trained upon a pair of images through using the proposed expanding strategy to create a training set and optimizing designed objective functions. The optimized GAN model would produce better coregistered images where changes can be easily spotted and then the change map can be presented through a comparison strategy using these generated images explicitly.
Compared to other deep learning-based methods, our method is less sensitive to the problem of unregistered images and makes most of the deep learning structure.
Experimental results on synthetic images and real data with many different scenes could demonstrate the effectiveness of the proposed approach.
\end{abstract}


\IEEEpeerreviewmaketitle

\section{Introduction}
\IEEEPARstart{C}{hange} detection in Earth Vision aims at generating the change map, that localizes the changed area in two satellite images which were taken at different times \cite{cd_1, cd_2}.
It is essential for many applications, such as urbanization monitoring \cite{app_1, app_2, app_3} and natural disaster analysing \cite{app_4, app_5}.
With multiple optical sensors available, i.e., Spot-5, Quickbird and Worldview, large amounts of satellite images with very high spatial resolution (VHR) can be obtained easily.
But it required much human intervention to identify changes in so many images. 
So change detection has arisen much more attention in recent years \cite{app_6, app_7}.

\begin{figure}[htbp]
    \centering
    \includegraphics[width=3in]{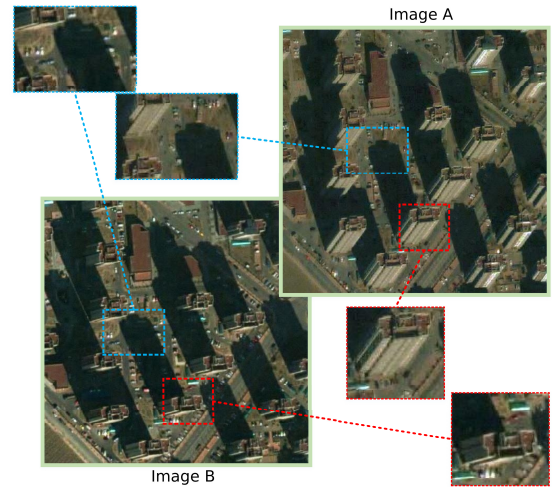}
    \caption{Example of unregistered paired images. The red rectangles enclose the same building that shows different appearances since the images were captured from different angles. The blue rectangles enclose the shadows which have different positions in two images.}
    \label{fig:uncor_illu}
\end{figure}

Supervised and unsupervised techniques are both commonly used in change detection.
Supervised methods usually transform change detection task to a classification that divides each pixel into two different classes \cite{camps2008kernel}.
However, since it is expensive to obtain large amount of annotated data that presents the regions of change directly, unsupervised methods are preferred over supervised ones in real cases.

Many of the unsupervised methods rely on the image coregistration that guarantees pixels, which have the same coordinates in two paired images, representing the same object.
\begin{figure*}[htbp]
    \centering
    \includegraphics[width=7.2in]{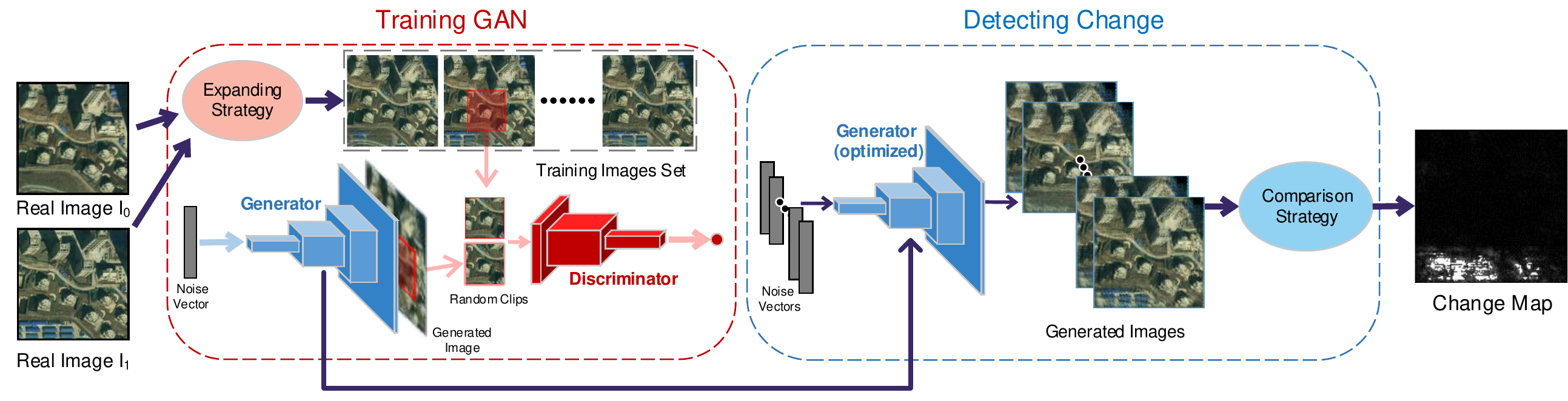}
    \caption{Proposed change detection framework. The original two images are firstly turned into the images set by proposed expanding strategy to train the GAN model. The generator is trained to produce images out of noise vectors $Z$ with the help of discriminator. After the model being optimized, the generator is capable of generating realistic coregistered images and a set of new images are created from randomly sampled noise vectors. Finally, the comparison strategy utilizes these generated images to get the change map.}
    \label{fig:procedure}
\end{figure*}
But nowadays, the satellite images describing the same scene may be taken by different sensors, from different angles or at different time in a day, which makes it difficult to obtain fully coregistered paired images, and thus giving rise to the inaccurate comparison between pixels. 
Fig. \ref{fig:uncor_illu} shows an example of unregistered scenes. The different appearances of the same buildings made the corresponding pixels apparently different and could also lead to different segmentation outputs. The corresponding shadows have similar appearances but different coordinates. 
Many of the unregistered objects are dispersed across the whole images and existing coregistration methods \cite{registration_1, registration_2} could not handle these cases perfectly even if more additional information is provided. Thus, many change detection methods suffered from such deficiency and the results of them could be affected heavily.

With convolutional neural network (CNN) shows strong capacity for extracting high-level features that are semantically rich \cite{feature_ext}, deep learning structures are applied into unsupervised methods to alleviate the effect of unregistered problems. The architectures with stacked convolution layers exploit the spatial context of pixels and could aggregate the features in an image into a low-dimensional vector. Analysing the high-level features could avoid pixel-level comparison.

However, there are still few unsupervised methods utilizing CNN architectures, and the limitations may come from the following aspects:
First, deep learning frameworks always serve as the intermediate procedure in unsupervised methods and complex calculations on their outputs are needed to produce results.
Second, to capture the features in images, the outputs of CNNs, which are termed feature hypervectors, are expected to well represent the images. But it is difficult to evaluate how changes on images would reflect on corresponding hypervectors since the absence of manual annotations.
Third, when pre-trained deep learning models are introduced, most of them are trained with the specific data with labels for classification or segmentation. Though deep learning-based feature extraction shows generality properties for transfer learning \cite{learningtransfer}, the quality of outputs is not guaranteed if a pre-trained model is used on a new data set without ground truth.
These points restrict the performance of CNNs for feature extraction in unsupervised methods. 
Therefore, frameworks for unsupervised tasks, in which deep learning architecture can achieve full potential, are demanded.

The existence of the unregistered pixels makes it difficult to locate the actual changed areas through the comparison on the unregistered images pairs or the feature hypervectors which are extracted by neural networks.
One straightforward solution is to employ the comparison on more and better corregistered images, thus alleviating the effect of unregistered pixels to the final results, and how to generate the desired images and how to get change map through these generated images then become the kernel.
Therefore, a special deep learning architecture, Generative Adversarial Network (GAN) \cite{gan}, is introduced to change detection procedure.
GAN actually consists of two individual neural networks that are termed generator and discriminator, respectively. Intuitively, the generator produces fake images taking uniformly sampled noise vectors as gradients and the discriminator judges images, giving high scores to real ones and low scores to fake ones. The training procedure helps the generator produce realistic images to fool the discriminator.
All generated images should be different and the differences are expected to be in changed objects, such as buildings, rivers or roads.
Meanwhile, the unchanged regions in generated images are expected to have alike appearances and the same positions, and this can be helpful for revealing actual changed areas.
A continuous high dimensional space for images is defined by the generator, and any images sampled from this space is useful for detecting changes, which ensures the deep learning architecture to be made full use of. The CNNs are served for generating and judging images, which means the limitations mentioned above could be avoided. 

The parameters in generator and discriminator are trained by optimizing respective objective functions. Objective function for discriminator combines the original Wassterin-GAN loss with an added term that shows good suitability for this task. 
Small clips, which are randomly selected from either generated images or training images, are fed to the discriminator and make it possible to produce coregistered images.
Moreover, to better indicate the desired images, a novel expanding strategy is proposed to create a training set out of the input images.
After acquiring the optimized model, images with difference in changed regions can be sampled and the final change map can be generated simply through the proposed comparison strategy.
Fig. \ref{fig:procedure} shows the framework of proposed method.

Contribution of this paper can be summarized as follows:
\begin{enumerate}
    \item Generative Adversarial Network is introduced in the unsupervised change detection task as an intermediate procedure to produce more images. It provides better coregistered images and makes it easy to reveal actual changed regions.
    \item New objective functions for GAN models are proposed to ensure no gradients vanishing or exploration during training process and thus ensure the feasibility of our method in theory.
    \item A novel expanding strategy is implemented to create more images for training data enhancement. This operation succeeds on indicating the desired images and helps to get proper GAN model in practice.
    \item A comparison strategy, which utilizes the generated images coming from optimized GAN model, is applied to produce the final change map.
\end{enumerate}

The rest of this paper is organized as follows. Section \uppercase\expandafter{\romannumeral2} presents the problem statement and describes the background of relative structures. The proposed change detection method is elaborated in Section \uppercase\expandafter{\romannumeral3}. Section \uppercase\expandafter{\romannumeral4} presents the experiments and analyses the results, some feature works are also discussed. In the end, this paper is concluded in Section  \uppercase\expandafter{\romannumeral5}.

\section{related works and background}

\subsection{Related Works}
Supervised techniques in change detection rely on the annotated data and the development of CNN has helped supervised change detection methods to obtain significant improvements.
Many different architectures have been explored.
\textsl{D. Peng et al.} \cite{unet++} proposed to use U-Net architecture to perform the direct learning of change map.
\textsl{M.A. Lebedev et al.} \cite{cgan_cd} utilized Conditional-GAN \cite{cgan} to recover the change maps out of the input images.
\textsl{Y. Zhan et al.} \cite{supervised_cd} proposed to extract features using Siamese CNN and obtain final change map in the supervised manner.

Unsupervised methods are usually based on the concept of Change Vector Analysis (CVA) \cite{cva}, which aims at rating each pixel the possibility of change. These methods can be further divided into two main types, pixel-based and object-based, according to different rating procedures.
Pixel-based methods treat pixels as independent individuals. PCA-Kmeans method \cite{pca-kmeans} applied Principal Component Analysis on the difference map acquired by subtracting corresponding pixels in paired images, then Kmeans-cluster were applied to divide pixels into different classes.
\textsl{C. Wu et al.} \cite{sfa} followed the idea of Slow Feature Analysis (SFA), finding a new space in which most of the pixels are invariant.
\textsl{F. Thonfeld et al.} \cite{rcva} introduced Robust CVA to mitigate the effects of pixel neighbourhood.
Pixel-based methods can have lots of data to analysis but could be affected by noise pixels or might lose much information between pixels.
Object-based change detection methods take spatial context information into account \cite{object-oriented2001} and various algorithms are developed to segment objects of interest \cite{object-based2008, object-based2012, object-based2018}. \textsl{L. Li et al.} \cite{ocva} proposed object-oriented CVA to compare each segment rather than single pixel. However, their results depend on the efficiency of segmentation process, which is still worthy of research.
In addition, some other types of change detection methods are developed, key points matching is one of them.
\textsl{G. Liu et al.} \cite{contrario} proposed to compute mappings between two images using SIFT-like \cite{sift} descriptors extraction and matching, then changed regions were indicated by mismatched descriptors.

With the popularity of deep learning in computer vision, many different deep learning architectures have been introduced to unsupervised change detection task.
\textsl{J. Liu et al.} \cite{bipartite} proposed an unsupervised method based on Restricted Boltzmann Machine to project the images into a low-dimension vector space, then change map can be generated by making the images' feature vectors much more similar.
\textsl{B. Du et al.} \cite{dsfa} proposed exploiting neural network to implement Slow Feature Analysis process.
\textsl{S. Saha et al.} \cite{unsupervisedCVA} used a pre-trained CNN to outcome feature vector, and a specific comparison and selection process was adopted.
\textsl{N. Lv et al.} \cite{autoencoders} introduced Contractive Autoencoder for extracting features from multiple images metrics. 

\subsection{Change Detection}
Let $I_0$ and $I_1$ be the two images used for change detection and $M$ be the change map where all pixel values indicate the possibility of change. The change detection task is formulated as:
\begin{equation}
    M=f(I_0,I_1), \label{1}
\end{equation}
where $f$ is the function that takes $I_0$ and $I_1$ as inputs and produces the change map. In unsupervised manners, $f$ always consists of $E$ and $f_A$. $E$ makes it easier to identify the difference between two images through some special operations, such as dimension transformation or feature extraction. $f_A$ represents the thresholds or clustering algorithms that are commonly used in unsupervised classification tasks. The formulation can be rewritten as:
\begin{equation}
    M=f_A[E(I_0, I_1)]. \label{2}
\end{equation}
Our method basically follows Equation \eqref{2}. Instead of extracting low dimensional feature vectors, $E$ is trained for producing realistic and different images with the same size of $I_0$ and $I_1$ .
Then $f_A$ uses these images to perform change detection process.
So, the final formulation is defined as follows:
\begin{equation}
    M=f_A[\mathop{\ominus}\limits_{i=1}^K E(z_i)], \label{3}
\end{equation}
where $z$ and $K$ denote the latent space vectors and the number of vectors used for generating change map, respectively. 

\begin{figure}
    \centering
    \includegraphics[width=3.2in]{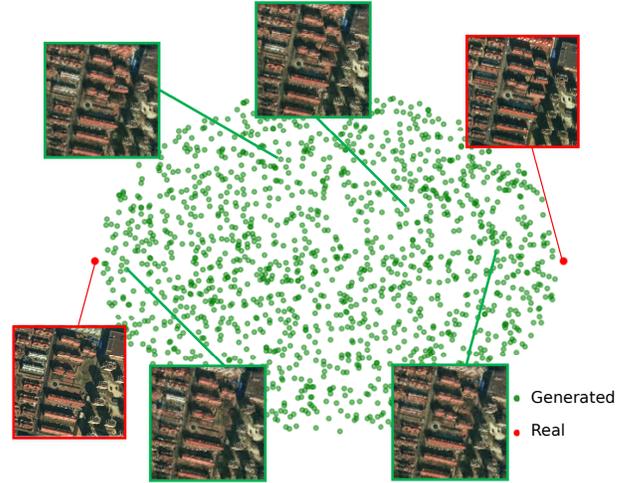}
    \caption{Illustration of the relationship between real images(red points) and generated images(green points) in 2D space view.}
    \label{fig:distribution}
\end{figure}

The relationship of changed regions in images could be understood from the perspective of data distribution.
$I_0$ and $I_1$ can be regarded as two separate points in a very high dimensional space. Its dimension is not as many as the image size $W \times H \times C$, since all pixel values have boundary and many objects in two satellite pictures are the same, we expect that variety only comes from the changed areas. Changes between two images can be understood as a straight line that links the two points, along which one point can move to the other. Change map is the representation of the length of this line and usually be called distance. Any distance can be measured if the space is defined and any points can be sampled from the space to which they belong.
Thus, what $E$ exactly does is to define the space, $\ominus$ gives a way to calculate the distance between the points sampled from the space, then these results are handled by $f_A$ to reveal the changed regions.
In this way, instead of extracting single feature vector, we find a feature space in which any point is a unique combination of features coming from the original pair. 

In Fig. \ref{fig:distribution} we show an example of the perspective in 2D coordinate. The red points are real images and green points represent the generated ones.
In the images space, all images describe the same scene but are different in details.
And the closer two points are, the more details in common are contained in these relative generated images \cite{model_space_3, model_space_1, model_space_2}.
This motivates us to utilize this space to alleviate the problem of unregistering. 
Images that show consistency in some small parts could be sampled from this space, while the differences still exist between images but much less than those in the original real images.
And some unregistered objects could be better coregistered in generated images.

\subsection{Generative Adversarial Network}
GAN consists of two CNN modules, a generator $G$ and a discriminator $D$. The generator learns a distribution $P_g$ over data $x$ via mapping the 1D vectors $z$, uniformly distributed input noise sampled from latent space $\mathcal{Z}$, to 2D images in the image space $\mathcal{X}$. The discriminator $D$ maps the images to a single value $D(\cdot)$ that represents the probability that the input was a real image comes from training data $\mathcal{X}$ or a fake image $G(z)$ generated by $G$.
$D$ and $G$ are optimized through playing the two-player minimax game and the objective functions are defined as \cite{gan}:
\begin{equation}
    L(D) = \max\limits_D \mathbb{E}_{x \sim P_{\mathcal{X}}}[\log D(x)]+\mathbb{E}_{x \sim P_g}[\log (1-D(x))], \label{4}
\end{equation}
\begin{equation}
    L(G) = \min\limits_G \mathbb{E}_{z \sim P_{\mathcal{Z}}}[1-\log D(G(z))], \label{5}
\end{equation}
where $P_{\mathcal{X}}$ and $P_{\mathcal{Z}}$ are the distribution of training data and noise, respectively. $D$ and $G$ will be optimized in turn. 

One of the most significant contribution of GAN is that it uses $D$, a function with CNN architecture, to define a measure of two distributions $P_g$ and $P_{\mathcal{X}}$. 
Divergence is widely used for measuring the distributions, and divergence $\mathcal{W}(P,Q)$ should have two properties: 
1) $\mathcal{W}(P,Q) \geq 0$ always holds, 
2) $\mathcal{W}(P,Q) = 0$ when $P = Q$. 
\textsl{I. Goodfellow et al.} \cite{gan} proved  when $D$ is optimized, $L(D)$ describes a divergence and $L(G)$ is trying to minimize the lower bound on the divergence.
Furthermore, the f-GAN \cite{f-gan} shows that slight modifications on objective \eqref{4} and \eqref{5} allows to lower bound on any desired f-divergence. 

\subsection{Wasserstein Generative Adversarial Network}
Though GANs have achieved great success at generating realistic and sharp looking images, they still remain difficult to train and suffer from training instability and gradients vanishing \cite{wgan-former}. \textsl{M. Arjovsky et al.} \cite{wgan} argued that when $P_{\mathcal{X}}$ and $P_g$ are supported on largely disjoint low-dimensional manifolds (which always happens in applications), Kullback-Leibler divergence, Jensen-Shannon divergence and many other distances between $P_{\mathcal{X}}$ and $P_g$ max out, so useful gradients are no longer provided for $G$. 
However, the 1-Wasserstein distance is still sensible in such situation and provides non-zero gradients. Thus, Wasserstein Generative Adversarial Network (WGAN) was proposed, and its objective functions were given:
\begin{equation}
    L(D) = \max\limits_{D, ||D||_L \leq 1}, \mathbb{E}_{x \sim P_{\mathcal{X}}}[D(x)]-\mathbb{E}_{x \sim P_{g}}[D(x)], \label{6}
\end{equation}
\begin{equation}
    L(G) = \min\limits_G -\mathbb{E}_{z \sim P_{\mathcal{Z}}}[D(G(z))], \label{7}
\end{equation}
where $||D||_L \leq 1$ indicates that the discriminator should satisfies the 1-Lipschitz constraint which is a special form of K-Lipschitz constraint, i.e., $|D(x_1)-D(x_2)| \leq K|x_1-x_2|$. The topic of enforcing Lipschitz constraints leaves open, many methods have been proposed \cite{wgan, wgan-gp, spectralnorm, gan-div}.

WGAN actually replaces common divergence with Wasserstein distance to measure two distributions. It shows that Equation \eqref{6} is still a divergence and Equation \eqref{7} makes the two distributions approximate to each other by minimizing the lower bound on the divergence. It also shows that no matter what data distribution is, WGAN will not suffer gradients vanishing theoretically if the discriminator $D$ is constrained in $||D||_L \leq 1$.

\section{methodology}
In this section, we will expound on the objective functions, the structure of GAN model, some other details of GAN training in the proposed method, and how to obtain the change map via generated images.
\subsection{Objective Functions}
Due to training data composing of very few images, the objective functions are based on the form of Wasserstein GAN to avoid gradients vanishing. Let $P_{\mathcal{I}}$, $P_g$ and $P_{\mathcal{Z}}$ be the distribution of training data, generated images and prior random noise. Inspired by GAN-QP \cite{ganqp}, the objective functions for the discriminator and the generator are defined as follow:
\begin{equation}
    \begin{aligned}
        L(D)=\max\limits_D \mathbb{E}_{(x_r,x_g)\sim (P_{\mathcal{I}},P_g)}&[D(x_r)-D(x_g)\\
        &-\lambda\frac{(D(x_r)-D(x_g))^2}{d(I_0,x_g)+d(I_1,x_g)}], \label{8}
    \end{aligned}
\end{equation}
\begin{equation}
    L(G)=\min\limits_G \mathbb{E}_{(x_r,z)\sim (P_{\mathcal{I}},P_{\mathcal{Z}})}[D(x_r)-D(G(z))], \label{9}
\end{equation}
where $x_r$ and $x_g$ refer to data points which are sampled from $P_{\mathcal{I}}$ and $P_g$ respectively, $I_0$ and $I_1$ denotes two images used for change detection, $\lambda$ is a hyperparameter that greater than zero and $d(\cdot)$ represents any distance, i.e., Euclidean distance.
Now, we will prove that Equation \eqref{8} is a divergence and $D$ satisfies Lipschitz constraint.

As have mentioned in Section \uppercase\expandafter{\romannumeral2}, Equation \eqref{8} can be regarded as a divergence if it is non-negative and equals to zero when $P_{\mathcal{I}}=P_g$. Thus, the proof consists of three parts.
\subsubsection{}
First, to prove $L(D) \geq 0$, we let $D(x) \equiv 0$ and we can get:
\begin{equation}
    L(D) \geq \mathbb{E}_{(x_r,x_g)\sim (P_{\mathcal{I}},P_g)}[0-\lambda\frac{0^2}{d(I_0,x_g)+d(I_1,x_g)}] = 0.
\end{equation}
Given that $L(D)$ selects the maximum of all possible $D$, so it is at least zero when we can always let $D(x) \equiv 0$.

\subsubsection{}
Secondly, if $P_{\mathcal{I}}=P_g$, we have:
\begin{equation}
    \begin{aligned}
        L(D)&=\max\limits_D \mathbb{E}_{(x_r,x_g)\sim (P_g,P_g)}[D(x_r)-D(x_g)\\
        &\quad\quad\quad\quad\quad\quad\quad\quad\quad -\lambda\frac{(D(x_r)-D(x_g))^2}{d(I_0,x_g)+d(I_1,x_g)}] \\
        &=\max\limits_D \mathbb{E}_{(x_r,x_g)\sim (P_g,P_g)}[-\lambda\frac{(D(x_r)-D(x_g))^2}{d(I_0,x_g)+d(I_1,x_g)}].
    \end{aligned}
\end{equation}
It is obvious that the max value is zero, so $L(D)=0$ when $P_{\mathcal{I}}=P_g$.

\subsubsection{}
Finally, if $P_{\mathcal{I}} \neq P_g$, let $p(x)$, $q(x)$ be the probability density function of $P_\mathcal{I}$ and $P_g$ respectively and
\begin{equation}
    D_0(x)=sign(p(x)-q(x)),
\end{equation}
then we have:
\begin{equation}
    \begin{aligned}
        L(D_0)&=\mathbb{E}_{x\sim P_\mathcal{I}}D_0(x)-\mathbb{E}_{x\sim P_g}D_0(x)\\
        &\quad\quad\quad\quad -\lambda\mathbb{E}_{(x_r,x_g)\sim (P_g,P_g)}\frac{(D_0(x_r)-D_0(x_g))^2}{d(I_0,x_g)+d(I_1,x_g)}] \\
        &=\int (p(x)-q(x))sign(p(x)-q(x))dx\\
        &\quad\quad\quad\quad -\lambda\mathbb{E}_{(x_r,x_g)\sim (P_g,P_g)}\frac{(D_0(x_r)-D_0(x_g))^2}{d(I_0,x_g)+d(I_1,x_g)}]\\
        &=t_1-t_2,
    \end{aligned}
\end{equation}
where $t_1$ and $t_2$ are two terms and both of them are greater than zero. When $t_1 \geq t_2$, then $L(D) \geq 0$. If $t_1 < t_2$, we can define:
\begin{equation}
    \begin{aligned}
        D(x) = \frac{t_1}{2t_2}D_0(x),
    \end{aligned}
\end{equation}
then
\begin{equation}
    \begin{aligned}
        &\mathbb{E}_{(x_r,x_g)\sim (P_{\mathcal{I}},P_g)}[D(x_r)-D(x_g)
        -\lambda\frac{(D(x_r)-D(x_g))^2}{d(I_0,x_g)+d(I_1,x_g)}]\\
        =&\frac{t_1}{2t_2}t_1-(\frac{t_1}{2t_2})^{2}t_2
        =\frac{t_1^2}{4t_2} > 0,
    \end{aligned}
\end{equation}
so $L(D) > 0$. 

The above three steps prove that the objective function for the discriminator (Equation \ref{8}) is a divergence. The proof utilizes many special cases of $D$ to construct the equations that satisfy required conditions. It is based on the fact that the training process would find the suitable form of $D$ and make the $L(D)$ reach its maximum. Thus, if a special case can be found, the optimized $D$ would at least be better than it and then satisfy the relative condition.

As have been proven in \cite{ganqp}, the optimum solution of Equation \eqref{8} satisfies:
\begin{equation}
    2\lambda\frac{D(x_r)-D(x_g)}{d(I_0,x_g)+d(I_1,x_g)}=\frac{p(x_r)q(x_g)-p(x_g)q(x_r)}{p(x_r)q(x_g)+p(x_g)q(x_r)}. \label{10}
\end{equation}
Note that $x_r$ is the sample of training data and can be $I_0$ or $I_1$, so it is obvious that:
\begin{equation}
    |D(x_r)-D(x_g)| \leq \frac{d(I_0,x_g)+d(I_1,x_g)}{2\lambda} \leq \frac{d(x_r, x_g)}{\lambda}. \label{Lip_equ}
\end{equation}
Thus, Equation \eqref{8} describes a reasonable divergence with the Lipschitz constraints enforced.

The objective function for generator is also a good measure of the distance between two distributions, and minimizing Equation \eqref{9} can push $P_g$ more close to $P_\mathcal{I}$ in $\mathcal{X}$.
After putting the optimum solution of Equation \eqref{10} into Equation \eqref{9}, the goal for generator to minimize can be obtained:
\begin{equation}
    \frac{1}{2\lambda}\mathbb{E}_{(x_r,x_g)\sim (P_\mathcal{I},P_g)}\frac{p(x_r)q(x_g)-p(x_g)q(x_r)}{p(x_r)q(x_g)+p(x_g)q(x_r)}\mathcal{D}(I_0,I_1,x_g), \label{11}
\end{equation}
\begin{equation}
    \mathcal{D}(I_0,I_1,x_g)=d(I_0,x_g)+d(I_1,x_g).
\end{equation}
To lower the value of \eqref{11}, the difference between $p(x)$ and $q(x)$ have to be lowered, which means the generator have to produce more realistic images.
$\mathcal{D}(I_0,I_1,x_g)$ serves as the weight, the further $x_g$ is away from $I_0$ and $I_1$, the greater the value will be. The generated data will get close to the training data to minimize $\mathcal{D}(I_0,I_1,x_g)$ and finally be settled in the middle space between $I_0$ and $I_1$, which is in line with what is expected.

Compared to the objective function, the denominator of the last term in Equation \ref{8} is replaced with the distance between generated images $x_g$ and two real images. 
This modification not only makes the calculating process efficiency since the real images are fixed, but also utilizes the feature of all images in training data which have almost the same scenes. 
It is an advantage when the term $d(I_0,x_g)+d(I_1,x_g)$ influences result of objective functions directly. 
This term encourages the generator to generate images with the same outfits since the term has to be lowered, and it also leaves room for variety since the partial combinations of $I_0$ and $I_1$ can also get good scores. 
The form of $L(D)$ restricts the generated images in a small proper space and makes it easy for convergence. 
Moreover, above demonstration proved that the objective functions still enforce the Lipschitz constraints.

\subsection{Support Expanding Strategy}
Support is the data can be seen in the training process.
The objective functions guarantee no gradients vanishing regardless of data distribution.
However, two images are too few to represent the whole space we want to find.
Given that the two images just like the loose boundary of images space and what we desire lay in between them, we propose to add more images as the support of training data set.
Instead of adding another real picture, the images are created by the linear combination of $I_0$ and $I_1$.

\begin{figure}[htbp]
    \centering
    \includegraphics[width=3in]{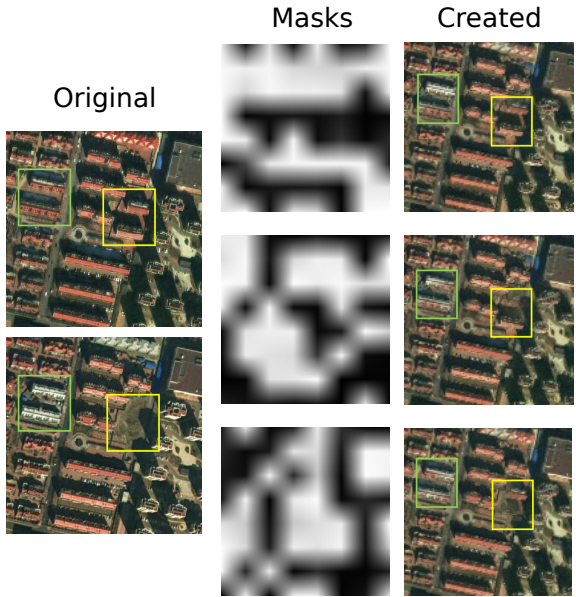}
    \caption{Examples of partial sampling masks. Green and yellow boxes enclose the major difference in two images and the created images show that masks can separate changed regions and reconstruct them efficiently.}
    \label{fig:mask_example}
\end{figure}

We suggest two ways to expand the training set.
\subsubsection{straight line sampling}
Let $\mathcal{I}=\{ i_0,i_1,\dots, i_n\}$ be the new training data set which consists of $n+1$ images, and
\begin{equation}
    i_k = \frac{k}{n+1}I_0+(1-\frac{k}{n+1})I_1. \label{creat_1}
\end{equation}
The new images are implicitly sampled uniformly along the straight line between the original pair of data, so we call it straight line sampling.
It is motivated by the space of interest which takes the images pair as boundary and sampling alone straight line is the simplest way to get points in that space.

\subsubsection{partial sampling}
\begin{figure}[htbp]
    \centering
    \begin{subfigure}[straight line sampling]{
        \includegraphics[width=3in]{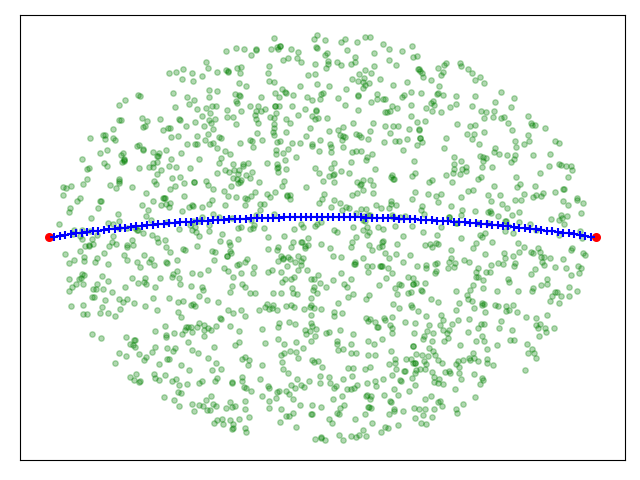}
    }
    \end{subfigure}
    
    \begin{subfigure}[partial sampling]{
        \includegraphics[width=3in]{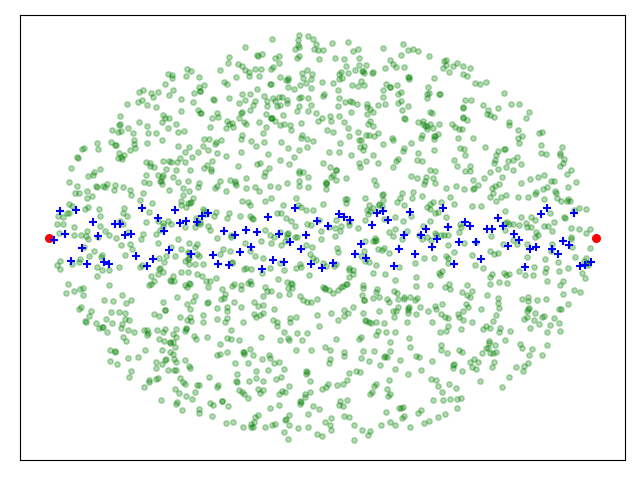}
    }
    \end{subfigure}
    \caption{The 2D space view for two support expanding strategies. Red points represent the real images while blue points represent the created ones.}
    \label{fig:support_expand}
\end{figure}

Let $\mathcal{M} = \{M_0, M_1, \dots, M_n\}$, and $M_k \in \mathcal{R}^{W \times H}$ be the single channel metric with the same size of $W \times H$ to the training data, each pixel in $M_k$ takes a value in the range of $[0,1]$, then the set can be expanded by
\begin{equation}
    i_k = M_k \cdot I_0+(1-M_k) \cdot I_1. \label{creat_2}
\end{equation}
$M_k$ just likes a mask that indicates how much weight of the original images should be in the new training data. Note that spatial context has to be taken into account, if all values in single mask $M_k$ are sampled individually, much of the information on the pixel neighborhoods would lose.
So we propose to form $M_k$ by up-sampling a tiny mask $m_k$ with a small size, i.e., $\frac{W}{16} \times \frac{H}{16}$, utilizing bilinear interpolation.
Besides, all values in $m_k$ are sampled from a non-uniform distribution where 1 and 0 have the highest probability. Some example masks after reshaping are shown in Fig. \ref{fig:mask_example}.
All images in $\mathcal{X}$ are generated by combining different part of $I_0$ and $I_1$ with different weight, so this method is called partial sampling and new data are the points sampled along lines with various directions.
Fig. \ref{fig:support_expand} shows some intuitive understanding of two sampling methods.

Straight line sampling is the subset of partial sampling when all pixels' values are the same in $M_k$. The former focuses on the changed regions in the whole image and realistic results can be generated at early stage, while the latter defines a much higher dimension and encourages generator to do more exploration.
Straight line sampling is recommended when the size of images is small. In such circumstance, the masks in partial sampling could not maintain the integrity in parts. But when the image size is large enough, i.e. $128 \times 128$, partial sampling is recommended to serve as the expanding strategy for creating training set.
$i_k$ in straight line sampling is determined by a single value $\frac{k}{n+1}$, so the degree of freedom\footnote{The degree of freedom refers to the number of parameters that control the system. In the expanding strategy, the degree of freedom is the number of parameters making the outputs different.} is 1, and that in partial sampling is controlled by the size of tiny mask $m_k$. If the length of input noise vectors is no less than the degree of freedom\footnote{When the model completes optimizing, all parameters are fixed. So the inputs control the outputs and the length of noise vectors is the degree of freedom of the optimized generator.}, it means the training data distribution $P_\mathcal{I}$ and generated data distribution $P_g$ can be aligned in image space $\mathcal{X}$.

\subsection{Network Architecture and Optimization}
The network in proposed method is based on the classical GAN structure given in \cite{gan}. The generator maps the latent space to the image space and the discriminator takes either generated images or images in training set as inputs, then produces single scalars.

The generator is a multi-layer perceptron where several transposed convolution layers are applied to up-sample low dimensional vectors to high dimensional ones. 
The distribution of noise vectors $P_Z$ in latent space would not matter and standard normal distribution $N(0,1)$ is always decided.
Here, the distributions with limited value range, i.e., uniform distribution $U[0,1]$, are suggested in this method while other distributions still work. It is motivated by the fact that the generated images are expected to be the fusion of $I_0$ and $I_1$ and the target distribution lays in the middle of these two points which can be regarded as the boundary. So the generated images are limited and choosing a distribution with limited value range makes it more in line with the target.

The discriminator encodes the input to a single scalar, so it encourages the low-frequency crispness. Though the support of training data has been expanded, the pictures in the set still describe the same scene. Considering that the changed regions are rare, the discriminator would focus on global structure of given images and return no information about details. Thus, the generator tends to produce almost the same images with no difference in regions.

\begin{algorithm}[t]
    \caption{Procedure of network training and change map generating}
    \label{algo-1}
    \hspace*{0.02in} {\bf Input:} 
        satellite images $I_0$, $I_1$, term weight $\lambda$, training batch size $m$, inference batch size $n$ and tiny image size $H, W$.\\
    \hspace*{0.02in} {\bf Output:}
        optimized parameters of the discriminator $\phi$ and the generator $\theta$
    \begin{algorithmic}[1]
    \State generate $\mathcal{I}=\{ i_0,i_1,\dots, i_m\}$ by Equation \eqref{creat_1} or \eqref{creat_2}.
    \State initialize discriminator parameters $\phi=\phi_0$ and generator parameters $\theta=\theta_0$
    \While{while $\theta$ has not converged}
        \State Sample training data $\{x_r^i\}_{i=1}^m \sim P_{\mathcal{I}}$, latent variable $\{z^i\}_{i=1}^m \sim P_\mathcal{Z}$, a random coordinate $(h,w)$
        \State $x_g \leftarrow G_\theta(z)$
        \State $\hat{x_g} \leftarrow x_g[h:h+H,w:w+W]$
        \State Calculate the loss of discriminator:
            \begin{equation}\nonumber
                \mathcal{L}(\phi)=[D_\phi(\hat{x_r})-D_\phi(\hat{x_g})
                -\lambda\frac{(D_\phi(\hat{x_r})-D_\phi(\hat{x_g}))^2}{d(I_0,x_g)+d(I_1,x_g)}] .
            \end{equation}
        \State Calculate the gradients and update $\phi$ utilizing gradient descent algorithm.
        \State Sample another batch of latent variable $\{z^i\}_{i=1}^m \sim P_\mathcal{Z}$
        \State Calculate the loss of generator:
            \begin{equation}\nonumber
                \mathcal{L}(\theta)=D_\phi(\hat{x_r})-D_\phi(G_\theta(z)[h:h+H,w:w+W]).
            \end{equation}
        \State Calculate the gradients and update $\theta$ utilizing gradient descent algorithm.
    \EndWhile
    \State Sample inference latent variable $\{z^i\}_{i=1}^n \sim P_\mathcal{Z}$
    \For{i=1,2,\dots,n}
        \State $x^i=G(z^i)$
    \EndFor
    \State Calculate the rough difference map:
        \begin{equation}\nonumber
            \Delta I = \frac{1}{n-1}\sum\limits_{i=2}^n threshold(|\frac{x^i}{max(x^i)}-\frac{x^1}{max(x^1)}|).
        \end{equation}
    \State Select max values over channels in $\Delta I$ to get the change map $CM$
    \State \Return Change map $CM$
    \end{algorithmic}
\end{algorithm}

This motivates us to restrict the discriminator to judge local structure by feeding a piece of clip that is cut from the image to the discriminator. This operation not only decreases the number of parameters in discriminator making it run faster, but also provides more scenes by dividing the original ones into many small parts. 
Given CNNs' bad performance on the invariance of pixel-level shifting \cite{shift-variance} and objects shifting a little can affect the outputs heavily, using small clips enforce the shift-invariance and helps the discriminator to be more robust. Meanwhile, the procedure of making discriminator shift-invariant can help the generator to produce better coregistered images, since the discriminator would be trained to ignore much pixel-level shifting. Then, the generator only has to produce images with objects that have common coordinate to get low losses from the discriminator.
Similar architecture were proposed in \cite{pix2pix, 2018patch} and they argued that such discriminators model images as a Markov random field \cite{mar_gan} and assume independence between pixels separated by more than a patch diameter \cite{pix2pix, mar_gan2, mar_gan3}.

To optimize the networks, the gradient descent step is performed on $D$ and $G$ in turn as suggested in \cite{gan} and the Adam solver \cite{Adam} is applied. 
The small clips will be uniformly extracted from either training data or generated images and then fed to the discriminator.
Thus, during the training process, the discriminator could get many different clips.

\subsection{Comparison Strategy for Change Map Generating}
When the optimized models are obtained, the generator learns the mapping $G(z)=z \rightarrow x$ from the latent space vectors to realistic images. 
These generated images have common coordinate in unchanged regions and show difference in changed regions. This is helpful to produce the change map by comparing the coregistered generated images directly and no more feature extraction or complex analyses are required.

We now propose a simple comparison strategy to reveal the changed regions.
Let $n$ be the pre-decided batch size and sample the latent variables batch $\{z^i\}_{i=1}^n \sim P_\mathcal{Z}$, then the realistic images would be obtained by $x^i=G(z^i)$, rough difference map can be calculated using
\begin{equation}
    \Delta I = \frac{1}{n-1}\sum\limits_{i=2}^n threshold(|\frac{x^i}{max(x^i)}-\frac{x^1}{max(x^1)}|),
\end{equation}
where $max(\cdot)$ gets the highest pixel value in $x^i$ and $threshold(\cdot)$ denotes the function that sets the values which are lower than 0.1 to 0. The rough difference map $\Delta I$ is a map with three channel and the large value in it means the high possibility for corresponding pixel being the changed one because the pixel has different value in most generated images. Accordingly, the highest value in each channel could be chosen for every pixel and $\Delta I$ can be turned to a single channel map. The threshold algorithm is employed to every difference map between two images and eliminates pixels with low possibility for being changed ones.

The generated images used for producing change map are sampled randomly and the randomness would not affect the results. The images come from the noise vectors that are uniformly sampled from latent space, and therefore could be regarded as being sampled uniformly from the image space. The sampled batch could well represent the distribution of generated space while the possibility of getting the batch with extreme distribution is approximately zero. On the other hand, more images can always be sampled to ensure the effectiveness of the change map.
The procedure of networks optimizing and producing change map is formally presented in Algorithm \ref{algo-1}.

\section{experiments and discussions}
In this section, we will present the experimental validation of the proposed method.
The data sets that are used will be detailed firstly, then the experimental setup.
To better illustrate how the method works, two experiments are implemented to employ our algorithm both on synthetic data and real satellite images. We focus not only on the change map results but also on the quality of the generated images.
Another comparative experiment will be presented to validate the effectiveness of the discriminator's structure. 

\subsection{Data Sets}
The proposed change detection framework was evaluated on synthetic images data set and real heterogeneous change detection data set Tianjin. The synthetic images contain geometric primitives that are generated randomly. The data images in Tianjin are further separated into three sets according to the scenes, rural scenes, urban scenes and scenes with no change. Some sets contain more than one pair of images. The Tianjin data set was acquired by WorldView-2 satellite sensor over an area of city TianJin, China, with a resolution of 0.4m/pixel and has a coverage of over 30 $km^2$. Two images in one pair were captured in year 2016 and 2017 in the same season, respectively. The objects in some scenes were shifted due to different incident angles. Some images pairs were annotated manually with the ArcGIS software to give the expectation of results. The original images from WorldView-2 have eight spectral bands. In our experiments, all images have been processed ahead and turned to common color images with three bands (R, G, B). The proposed methods only utilized the three-bands images as input and no more other information or data were needed.

\subsection{Experimental Setup}
Several parameters or functions need to be set in our method, they will be described here in detail.
\subsubsection{$\lambda$ in Discriminator's Objective Function}
$\lambda$ controls the value of the newly added item in Equation \eqref{8} and decides the value $K$ of K-Lipschitz constraint as shown in Equation \eqref{Lip_equ}.
We set $\lambda=0.2$ to make the item's value have an equal magnitude as term $D(x_r)-D(x_g)$
in $L(D)$.
However, other values, i.e., $\lambda=0.15$ or $\lambda=0.25$ also work and will not affect the results too much.
\subsubsection{Image Size}
There are two kinds of size in proposed method, the size of images in training data and size of input clips for the discriminator.
All images in training data set $\mathcal{I}$ are of size $128\times128\times3$. Satellite images in this size are large enough to display changed objects. Images with larger size won't present better results and make the generator require more parameters and more time for training.
Tiny clips that fed to the discriminator are of size $64\times64\times3$ and they are captured randomly from training data or generated images.
\subsubsection{Distance Function $d(\cdot)$}
It is an important function that influences the loss value of discriminator. L2-distance, also known as MSE distance is employed in the following experiments.
The function $d(\cdot)$ takes the two images as inputs and computes the sum of square error of each pixel. One advantage is that L2-distance penalizes pixel pairs which have great difference, so the effect of some pixel-level noise can be weakened. But the choice of $d(\cdot)$ remains open and any measure of similarity of two pictures can be tried.
\subsubsection{Structure of CNN}
Deconvolutional layers and convolutional layers are employed to implement the generator and discriminator, in which the convoluntional kernels are all of size $4\times4$. The size of latent space vectors is $64\times1$. The stride for each kernel is set to 2 to reduce the size of inputs instead of applying pooling layer. The batch normalization \cite{batch_norm} is also applied and batch size is set to 64. Halt is another open issue for GANs' training procedure and we found the outputs after 500 iterations are good enough. Furthermore, the number of steps in every training epoch is set to 50. The training rates of generator and discriminator are both set to 1e-4, and the batch size in every training step is set to 64. To ensure the variance in training data, the data set is expanded to contain 3200 images, so that every training sample in one epoch is unique.

Some other unsupervised change detection methods, like CVA, PCA-Kmeans \cite{pca-kmeans}, PCANet \cite{pca-net}, BDNN \cite{bipartite} and DSFA \cite{dsfa}, are implemented in experiments for comparison.
The change maps would be finally evaluated based on six criteria, including precision–recall (PR) curve, the Receiver Operating Characteristic (ROC) curve, overall accuracy (OA), precision score, Kappa coefficient and F1 score.
Change map gives the change intensity for each pixel and will be separated into two parts using different thresholds ranging from 0 to 1 to generate a binary map. This binary map then is compared with annotation to obtain the precision, the true positive rate and the false positive rate on which these criteria are based. PR and ROC curves are based on the single change map and a good change map's PR curve is close to the top-right corner while ROC curve is close to the top-left corner. 

The quality of generated images is also of great concern, and we choose Fréchet Inception Distance (FID) \cite{FID} to measure the distance between training data distribution and generated data distribution. FID utilizes a pre-trained Inception Net to extract features and outputs a final score. The smaller the score is, the better the generated images are.

\begin{figure}[t!]
    \centering
    \begin{subfigure}{
        \includegraphics[width=0.82in]{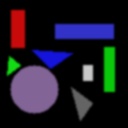}
    }
    \end{subfigure}
    \hspace{-0.55cm}
     \begin{subfigure}{
        \includegraphics[width=0.82in]{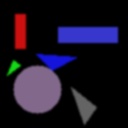}
    }
    \end{subfigure}
    \hspace{-0.55cm}
    \begin{subfigure}{
        \includegraphics[width=0.82in]{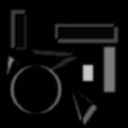}
    }
    \end{subfigure}
    \hspace{-0.55cm}
    \begin{subfigure}{
        \includegraphics[width=0.82in]{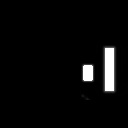}
    }
    \end{subfigure}
    \caption{Synthetic data set, from left to right: input image A, input image B, absolute change map, reference change map. The absolute change map was generated by subtracting pixels in A from corresponding pixels in B, so the influence of objects shifts was shown clearly in it.}
    \label{fig:syn_data}
\end{figure}

\begin{figure}[t!]
    \centering
    \begin{subfigure}{
        \includegraphics[width=0.82in]{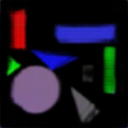}
    }
    \end{subfigure}
    \hspace{-0.55cm}
     \begin{subfigure}{
        \includegraphics[width=0.82in]{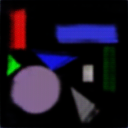}
    }
    \end{subfigure}
    \hspace{-0.55cm}
    \begin{subfigure}{
        \includegraphics[width=0.82in]{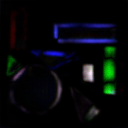}
    }
    \end{subfigure}
    \hspace{-0.55cm}
    \begin{subfigure}{
        \includegraphics[width=0.82in]{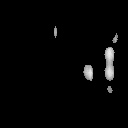}
    }
    \end{subfigure}
    \caption{Generated images and results of algorithm, from left to right: generated image $G_1$, generated image $G_2$, $\Delta I$, change map. The gray rectangle and green rectangle were presented in $G_1$ and $G_2$ respectively, other primitives were recovered and remained almost the same. The change map reveals the changed regions}
    \label{fig:syn_gen}
\end{figure}

\subsection{Experiment on Synthetic Images}

In the first experiment, our method was tested on the synthetic data set. Synthetic data set consists of two images in the size of $128\times128\times3$. Both images have black background and several nonintersecting geometric primitives (rectangle, round, and triangle). Some geometric primitives presented only in one image represent the changed objects. To simulate the situation that images are taken by different devices or from different angles, the color and the size of corresponding objects were not the same.
The value difference of single value for each pixel was 10, but this difference may be hard to be spotted by human sight. The pixel-level shifts for each object were in range [0, 5] and the shifts occurred in horizontal and vertical directions. Two images were smoothed by a Gaussian filter with $\sigma=0$. Fig. \ref{fig:syn_data} shows the synthetic images, absolute difference map and reference change map.

We employed partial sampling to expand support, and after 500 training iterations, we finished optimizing GAN model then utilized it to generate change map. Fig. \ref{fig:syn_gen} shows two random sampled generated images, $\Delta I$ and change map result. As we can see, most of the changed regions are detected and the effect of objects shifts is mitigated, though some false detection still exists. Simple comparison strategy was applied and pixels with value less than half of the maximum pixel value were dropped. The way the training set was expanded made it hard to keep edges of primitives and the generator could not recover the edges perfectly. As a consequence, the regions in change map look irregular
and fragmentary, but they still point out where changes happened.

The value of FID for synthetic data set was 82.13, precision and recall values in this experiment were 0.838 and 0.761 respectively when threshold was set to 0.020.

\subsection{Experiments on Real Heterogeneous Images}

\begin{figure}[t!]
    \centering
    \begin{subfigure}[]{
        \includegraphics[width=1in]{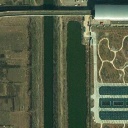}
    }
    \end{subfigure}
    \hspace{-0.55cm}
    \begin{subfigure}[]{
        \includegraphics[width=1in]{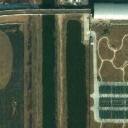}
    }
    \end{subfigure}
    \hspace{-0.55cm}
    \begin{subfigure}[]{
        \includegraphics[width=1in]{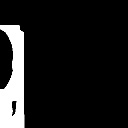}
    }
    \end{subfigure}
    \caption{Rural scene, (a) image acquired in 2016, (b) image acquired in 2017, (c) manual annotation.}
    \label{fig:o_864}
\end{figure}

\begin{figure}[t!]
    \centering
    \begin{subfigure}[]{
        \includegraphics[width=1in]{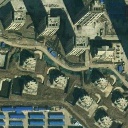}
    }
    \end{subfigure}
    \hspace{-0.55cm}
    \begin{subfigure}[]{
        \includegraphics[width=1in]{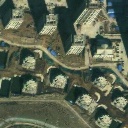}
    }
    \end{subfigure}
    \hspace{-0.55cm}
    \begin{subfigure}[]{
        \includegraphics[width=1in]{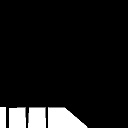}
    }
    \end{subfigure}
    \caption{Urban scene, (a) image acquired in 2016, (b) image acquired in 2017, (c) manual annotation.}
    \label{fig:o_187}
\end{figure}

\begin{figure}[t!]
    \centering
    \begin{subfigure}[]{
        \includegraphics[width=1in]{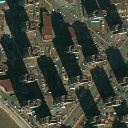}
    }
    \end{subfigure}
    \hspace{-0.55cm}
    \begin{subfigure}[]{
        \includegraphics[width=1in]{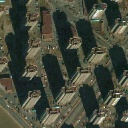}
    }
    \end{subfigure}
    \hspace{-0.55cm}
    \begin{subfigure}[]{
        \includegraphics[width=1in]{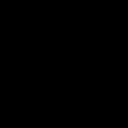}
    }
    \end{subfigure}
    \caption{Scene with no changes, (a) image acquired in 2016, (b) image acquired in 2017, (c) manual annotation.}
    \label{fig:o_555}
\end{figure}

Real heterogeneous images in the data set can be divided into three parts.
The first part of data describes the rural area, there are many fields and small low buildings in the pictures. 
Then we got the second part, images describing urban area suffer from unregistered problem caused by tall buildings and their shadows. 
The third part includes special image pairs with no changed regions, so their change maps are blank and no PR or ROC curves can be plotted because the true positive rate is always 0. We testified proposed method on above three types of data.

\begin{figure*}[t!]
    \centering
    \begin{subfigure}[]{
        \includegraphics[width=1in]{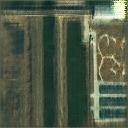}
    }
    \end{subfigure}
    \hspace{-0.55cm}
    \begin{subfigure}[]{
        \includegraphics[width=1in]{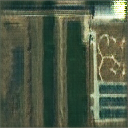}
    }
    \end{subfigure}
    \hspace{-0.55cm}
    \begin{subfigure}[]{
        \includegraphics[width=1in]{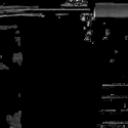}
    }
    \end{subfigure}
    \hspace{-0.55cm}
    \begin{subfigure}[]{
        \includegraphics[width=1in]{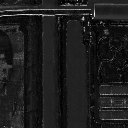}
    }
    \end{subfigure}
    \hspace{-0.55cm}
    \begin{subfigure}[]{
        \includegraphics[width=1in]{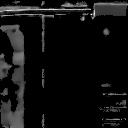}
    }
    \end{subfigure}
    \hspace{-0.55cm}
    \begin{subfigure}[]{
        \includegraphics[width=1in]{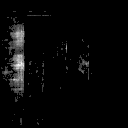}
    }
    \end{subfigure}
    \caption{Change maps given by different methods on rural scene: (a) and (b) two samples of generated images, (c) PCA-Kmeans, (d) DSFA, (e) PCA-Net, (f) proposed method.}
    \label{fig:864}
\end{figure*}

\begin{figure*}[t!]
    \centering
    \begin{subfigure}[]{
        \includegraphics[width=1in]{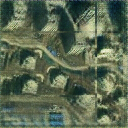}
    }
    \end{subfigure}
    \hspace{-0.55cm}
    \begin{subfigure}[]{
        \includegraphics[width=1in]{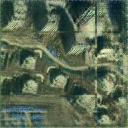}
    }
    \end{subfigure}
    \hspace{-0.55cm}
    \begin{subfigure}[]{
        \includegraphics[width=1in]{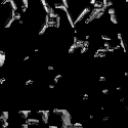}
    }
    \end{subfigure}
    \hspace{-0.55cm}
    \begin{subfigure}[]{
        \includegraphics[width=1in]{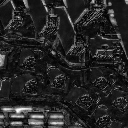}
    }
    \end{subfigure}
    \hspace{-0.55cm}
    \begin{subfigure}[]{
        \includegraphics[width=1in]{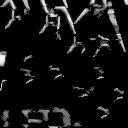}
    }
    \end{subfigure}
    \hspace{-0.55cm}
    \begin{subfigure}[]{
        \includegraphics[width=1in]{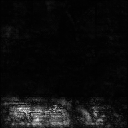}
    }
    \end{subfigure}
    \caption{Change maps given by different methods on urban scene: (a) and (b) two samples of generated images, (c) PCA-Kmeans, (d) DSFA, (e) PCA-Net, (f) proposed method.}
    \label{fig:187}
\end{figure*}

\begin{figure*}[t!]
    \centering
    \begin{subfigure}[]{
        \includegraphics[width=1in]{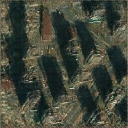}
    }
    \end{subfigure}
    \hspace{-0.55cm}
    \begin{subfigure}[]{
        \includegraphics[width=1in]{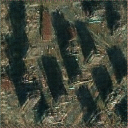}
    }
    \end{subfigure}
    \hspace{-0.55cm}
    \begin{subfigure}[]{
        \includegraphics[width=1in]{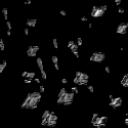}
    }
    \end{subfigure}
    \hspace{-0.55cm}
    \begin{subfigure}[]{
        \includegraphics[width=1in]{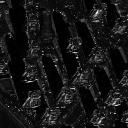}
    }
    \end{subfigure}
    \hspace{-0.55cm}
    \begin{subfigure}[]{
        \includegraphics[width=1in]{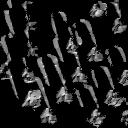}
    }
    \end{subfigure}
    \hspace{-0.55cm}
    \begin{subfigure}[]{
        \includegraphics[width=1in]{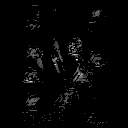}
    }
    \end{subfigure}
    \caption{Change maps given by different methods on scene with no changes: (a) and (b) two samples of generated images, (c) PCA-Kmeans, (d) DSFA, (e) PCA-Net, (f) proposed method.}
    \label{fig:555}
\end{figure*}

\subsubsection{Experiment on Rural Scene}
Images with fields, water and low buildings can be regarded as roughly coregistered image pairs because those objects have similar appearances, but some pixel-level shifts still exist. Fig. \ref{fig:o_864} gives the images used in the experiment along with manual annotation of change. The regions of interest are in left of this scene, and change map generated by different methods are presented in Fig. \ref{fig:864} (c), (d), (e) and (f). The shifts of objects in upper part affected the results, especially pixel-based method like DSFA, and the false detected pixels made it difficult to tell where true changed regions were from the change map. Though the proposed method did not give out the whole changed area, it was robust to influence of such shifts. Two examples of generated images are shown in Fig. \ref{fig:864} (a), (b), and they showed obvious difference in annotated regions. Meanwhile, the objects at upper and right part in generated images all had alike appearance and consistent coordinates. Fig. \ref{fig:curv} (a), (c) show the PR and ROC curves, respectively. 

The change detection results over all images with rural scenes are presented in Table \ref{tab:result_rural} with the best values of each evaluation criteria highlighted with bold. All methods achieve high scores on overall accuracy but had low values on Kappa coefficient and F1 score. PCA-Kmeans and DSFA were suffered from the existence of many pixels that were falsely detected and the relative scores would decrease when false positive rate (FPR) matters a lot.
It can be observed that the proposed method achieved better result compared to other methods and outperformed them due to the advantage on handling pixel-level shifts.

\subsubsection{Experiment on Urban Scene}
\begin{figure}[t!]
    \centering
    \includegraphics[width=3.5in]{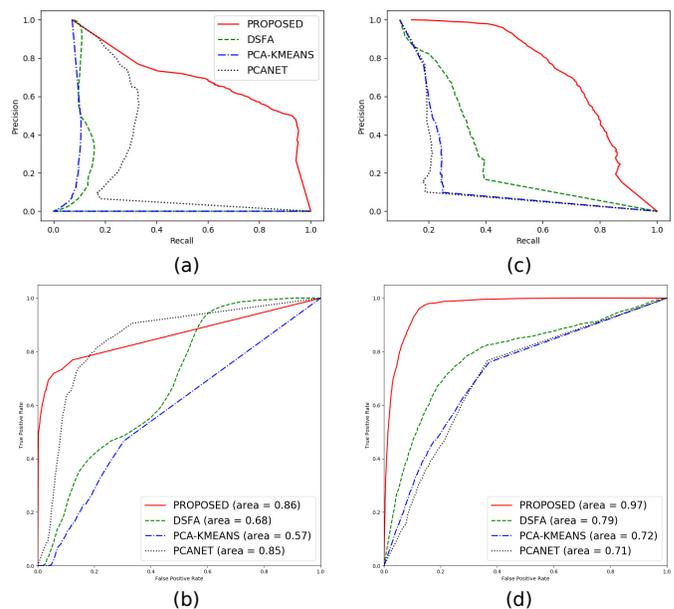}
    \caption{PR and ROC curves for the experiments on rural and urban scenes. (a) PR curve of rural scene experiment, (b) PR curve of urban scene experiment, (c) ROC curve of rural scene experiment, (d) ROC curve of urban scene experiment.}
    \label{fig:curv}
\end{figure}
It is much more difficult to detect changes on images describing urban area, not only for the unregistered problem, but also because these images have complex texture which makes it hard for generator to produce realistic images. A typical pair of images and their annotations are given in Fig. \ref{fig:o_187}. 
As shown in images, the change comes from blue buildings in bottom left, however, some parts of ground in the upper half of Fig. \ref{fig:o_187} (a) are covered by shadows while those in (b) are not. The values of pixels that correspond to shadows and ground differ greatly. 
Fig. \ref{fig:187} shows the results of this experiment, and the change maps given by other methods suffered a lot from unregistered pixels, although they still caught parts of the changed regions. 
The result of proposed method performed good on ignoring those unregistered parts and giving exact regions of interest. From the two examples in Fig. \ref{fig:187} (a) and (b), we can see that areas other than the annotated ones just remained the same. The quality of generated images was not good enough, and there were tearing in them because it is difficult for CNN to recover edge areas. But the generated images still recovered the position and outfit of main objects in the scene. 
PR and ROC curves are shown in Fig. \ref{fig:curv} (b), (d) and the results of urban scene data are given in Table \ref{tab:result_urban}. Comparing with the results of rural scene, nearly all scores became worse since the complex scenes introduced more unregistered areas. The proposed method gives better performance on alleviating such effect and gets obvious improvements in terms of Kappa coefficient and F1 score.

\begin{table}[t!]
    \renewcommand{\arraystretch}{1.3}
    \caption{Change detection results of Rural Scene Data}
    \label{tab:result_rural}
    \centering
    \begin{tabular}{c||c|c|c|c}
    \hline
     & \bfseries OA & \bfseries Pre & \bfseries Kappa & \bfseries F1\\
    \hline\hline
    \bfseries PCA-Kmeans & 0.8019 & 0.7751 & 0.5731 & 0.6135\\
    \hline
    \bfseries DSFA & 0.7607 & 0.7842 & 0.5908 & 0.6237\\
    \hline
    \bfseries PCA-Net & 0.8846 & 0.8154 & 0.6718 & 0.7375\\
    \hline
    \bfseries Proposed & \bfseries 0.9201 & \bfseries 0.8266 & \bfseries 0.7003 & \bfseries 0.7503\\
    \hline
    \end{tabular}
\end{table}

\begin{table}[t!]
    \renewcommand{\arraystretch}{1.3}
    \caption{Change detection results of Urban Scene Data}
    \label{tab:result_urban}
    \centering
    \begin{tabular}{c||c|c|c|c}
    \hline
     & \bfseries OA & \bfseries Pre & \bfseries Kappa & \bfseries F1\\
    \hline\hline
    \bfseries PCA-Kmeans & 0.7931 & 0.6930 & 0.6631 & 0.7172\\
    \hline
    \bfseries DSFA & 0.7586 & 0.7075 & 0.6215 & 0.6916\\
    \hline
    \bfseries PCA-Net & 0.7889 & 0.7101 & 0.6394 & 0.7324\\
    \hline
    \bfseries Proposed & \bfseries 0.8742 & \bfseries 0.7551 & \bfseries 0.7093 & \bfseries 0.7692\\
    \hline
    \end{tabular}
\end{table}

\begin{table}[t!]
    \renewcommand{\arraystretch}{1.3}
    \caption{Change detection results of No Changed Scene Data}
    \label{tab:result_nochange}
    \centering
    \begin{tabular}{c||c|c|c|c}
    \hline
     & \bfseries OA & \bfseries Pre & \bfseries Kappa & \bfseries F1\\
    \hline\hline
    \bfseries PCA-Kmeans & 0.8300 & - & - & -\\
    \hline
    \bfseries DSFA & 0.7175 & - & - & -\\
    \hline
    \bfseries PCA-Net & 0.6908 & - & - & -\\
    \hline
    \bfseries Proposed & \bfseries 0.9351 & - & - & -\\
    \hline
    \end{tabular}
\end{table}

\subsubsection{Experiment on No Changed Scene}
Unsupervised change detection methods cannot judge whether there are changes because no labels are provided, and the results always show the pixels that are most likely to be classified as changed ones. 
\begin{figure*}[htbp]
    \centering
    \includegraphics[width=7in]{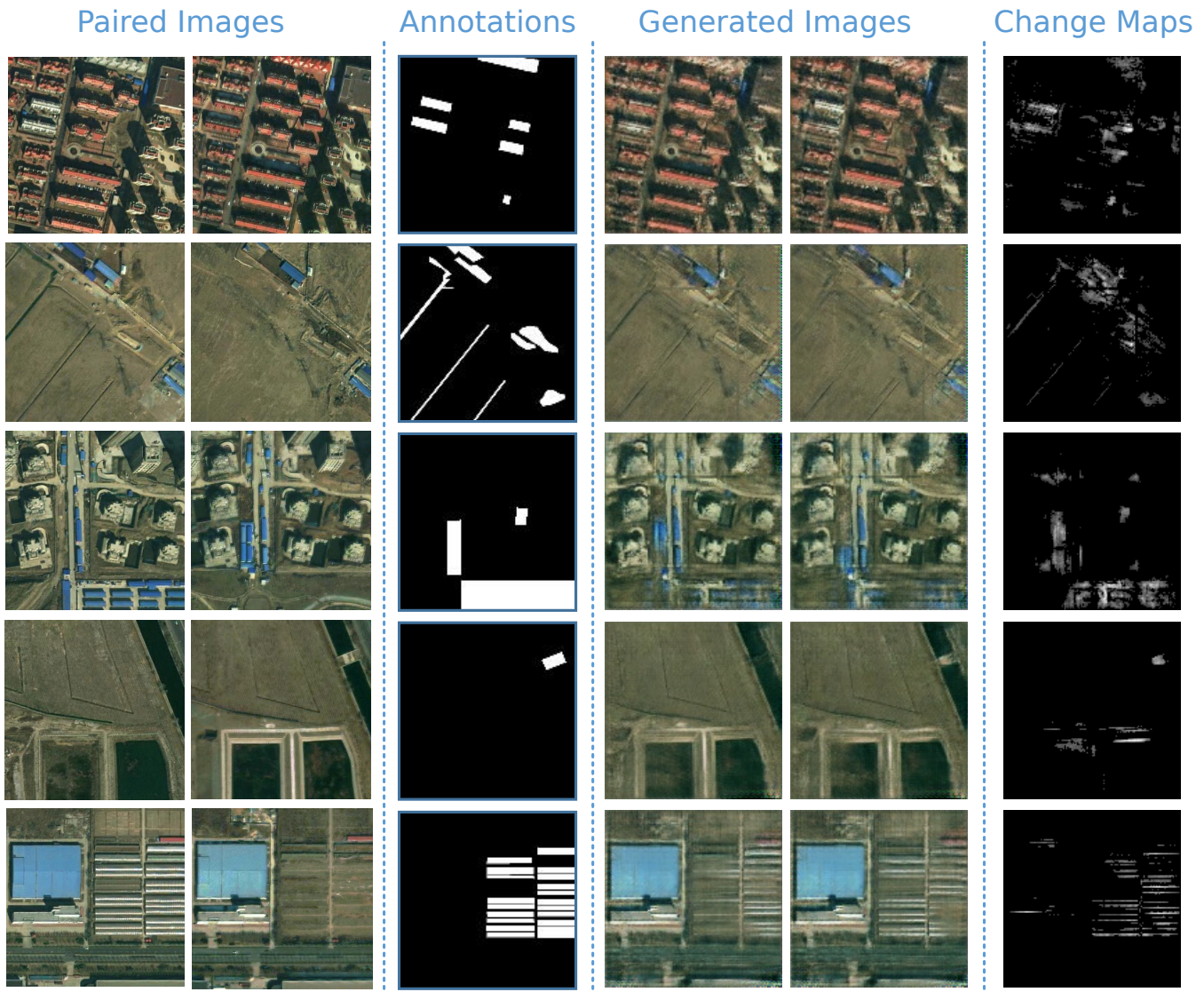}
    \caption{Satellite images change detection collection. More results of various scenes are presented here.}
    \label{fig:many_exam}
\end{figure*}
Fig. \ref{fig:o_555} gives two images containing the same buildings and there are no changed regions. According to the discussion in Section \uppercase\expandafter{\romannumeral1}, different shooting angle and time lead to two different satellite images, and as fewer pixels were expected in final change map as possible. Compared to the change maps of other methods in Fig. \ref{fig:555}, the result of proposed algorithm has the fewest false positive pixels. 
Two generated images shown in Fig. \ref{fig:555} (a) and (b) are blurry and the blur is formed by putting the corresponding same objects into one images. The contours of tall buildings with different appearances occurred in same places, and shadows had common shapes and coordinates with difference in brightness. The differences between any two sampled images were various, so values of most pixels in $\Delta I$ were small and the corresponding pixels would be wiped by setting the threshold. Table \ref{tab:result_nochange} presents the criteria on all data with no changed scene. Only overall accuracy could be calculated since there are no true positive pixels. The proposed method achieved the highest score, which means that it had the fewest false detected pixels. This experiment shows the advantage of utilizing much more images than two to decide changed pixels.

From the three experiments above, we demonstrated that the proposed method achieved as good performance on real satellite images as many other methods did.
Furthermore, the proposed method is robust to objects shifts and can mitigate the impact of unregistered problems caused by different appearances of the same objects. Here, more results on various scenes are presented in Fig. \ref{fig:many_exam}

\subsection{Discussion on Discriminator's Structure}
\begin{figure}[t!]
    \centering
    \begin{subfigure}{
        \includegraphics[width=1.1in]{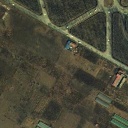}
    }
    \end{subfigure}
    \hspace{-0.55cm}
    \begin{subfigure}{
        \includegraphics[width=1.1in]{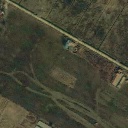}
    }
    \end{subfigure}
    \hspace{-0.55cm}
    \begin{subfigure}{
        \includegraphics[width=1.1in]{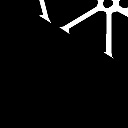}
    }
    \end{subfigure}
    \caption{Two images for the experiments of discriminator's structure and manual annotation.}
    \label{fig:0}
\end{figure}

In this section, we focused on the structure of discriminator, and it assists to train a good generator which is crucial to the whole procedure. 
As mentioned in Section \uppercase\expandafter{\romannumeral3}.C, the discriminator only takes part of images as input, so discriminator's structure changes when the size of input images becomes different. Following experiment gives the comparison on the qualities of images generated by different discriminators. 
The paired images shown in Fig. \ref{fig:0} were used in this experiment and both images are of size $128\times128$.

\begin{figure}[t!]
    \centering
    \begin{subfigure}{
        \includegraphics[width=1in]{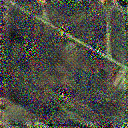}
    }
    \end{subfigure}
    \hspace{-0.4cm}
    \begin{subfigure}{
        \includegraphics[width=1in]{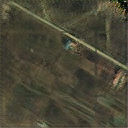}
    }
    \end{subfigure}
    \hspace{-0.4cm}
    \begin{subfigure}{
        \includegraphics[width=1in]{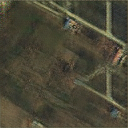}
    }
    \end{subfigure}
    
    \vspace{-0.3cm}
    \begin{subfigure}{
        \includegraphics[width=1in]{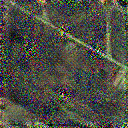}
    }
    \end{subfigure}
    \hspace{-0.4cm}
    \begin{subfigure}{
        \includegraphics[width=1in]{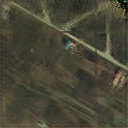}
    }
    \end{subfigure}
    \hspace{-0.4cm}
    \begin{subfigure}{
        \includegraphics[width=1in]{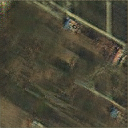}
    }
    \end{subfigure}
    
    \vspace{-0.3cm}
    \begin{subfigure}{
        \includegraphics[width=1in]{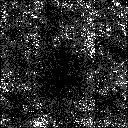}
    }
    \end{subfigure}
    \hspace{-0.4cm}
    \begin{subfigure}{
        \includegraphics[width=1in]{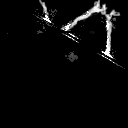}
    }
    \end{subfigure}
    \hspace{-0.4cm}
    \begin{subfigure}{
        \includegraphics[width=1in]{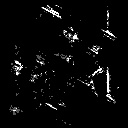}
    }
    \end{subfigure}
    \caption{Generated samples and change detection results from different discriminator: left - DIS-128, middle - DIS-64, right - DIS-32.}
    \label{fig:diff_dis_size}
\end{figure}

Three sizes ($128\times128$, $64\times64$, $32\times32$) were tested and three models with different structures of the discriminator, which were termed DIS-128, DIS-64 and DIS-32, respectively, were obtained after being trained for 600 iterations. From the results (Fig. \ref{fig:diff_dis_size}), we can see the influence of different structure choices. 
The first column in Fig. \ref{fig:diff_dis_size} contains two samples got from DIS-128. The images have many noise pixels and it is hard to tell the difference between them. 
Considering that the input images are of full size and the discriminator just encodes inputs into single values, much information would be ignored during the encoding process. So the discriminator was mostly influenced by the overall scenes and there was no information back for generator to handle the details.

Two images in the second column are results of DIS-64 and change map, which is the recommended structure in the experiment. Compared to DIS-128, it has small varied image clips for discriminator and information for details can be cared through back-propagation.
The last column shows the outputs of DIS-32. Interestingly, they got details and variance but the difference was in the wrong place. Smaller clips were selected randomly then fed to the discriminator in DIS-32 and no information about position was provided, so discriminator never knew which part it was criticizing, then generator could draw this part in anywhere it liked. This problem did not happen in DIS-64, because the size $64\times64$ made nearly all clips have overlapping part which could indicated the positions actually. When size of clips decreased, many clips became uncorrelated and indications of position disappeared, the absence of clips' position resulted in the displacement of many visual features.

The FID scores (Table \ref{tab:FID_scores}) demonstrated that DIS-128 has the worst result which is in line with the intuition. DIS-64 and DIS-32 have better scores since they all have good details and FID do not implement the criterion to tell the displacement in images. Figure \ref{fig:diff_dis_size} also presents the change maps generated by implementing different discriminators. The result of DIS-128 contained much noise and did not locate the actual changed areas. The changed areas in the result of DIS-32 showed similar shapes with the annotation but were not in accurate positions. The results of DIS-32 was the most reasonable one and the score of all criteria in Table \ref{tab:FID_scores} also verified the effect of different discriminators on the results.

\begin{table}[t!]
    \renewcommand{\arraystretch}{1.3}
    \caption{FID scores and change detection results of different discriminators}
    \label{tab:FID_scores}
    \centering
    \begin{tabular}{c||c|c|c}
    \hline
     & \bfseries DIS-128 & \bfseries DIS-64 & \bfseries DIS-32\\
    \hline\hline
    \bfseries FID & 284.49 & 163.37 & 167.11\\
    \hline
    \bfseries OA & 0.5582 & 0.9471 & 0.8990\\
    \hline
    \bfseries Pre & 0.2672 & 0.6731 & 0.3751\\
    \hline
    \bfseries Kappa & 0.1328 & 0.5953 & 0.1279\\
    \hline
    \bfseries F1 & 0.2209 & 0.6226 & 0.1810\\
    \hline
    \end{tabular}
\end{table}

The results above show that a balance should be found when choosing the structure of $D$. The size of input clips needs to be smaller than that of original images, but it cannot be too small or the lack of position information makes it difficult for generator to recover objects in right places.So the size $64\times64$ was suggested in proposed method. 

The results also motivated us to come up a way utilizing the coordinate of clips explicitly. Just like many Objects Detection Networks did for regression of objects' positions, an auxiliary learning task \cite{aux-learn} could be added into discriminator to learn the coordinates. However, a new learning task means a new term in objective function and objective function need to be re-designed to satisfy Lipschitz constraints again which is hard to work out.

\subsection{Runtime Analysis}
In this section, we would discuss the computing time of the proposed framework. Most of the time came from the training process. However, it is not very time consuming even if multiple convolution layers were implemented. The comparison on runtime of PCA-Kmeans, DSFA, \textit{Contrario} framework, PCA-Net and proposed method are presented in Table \ref{tab:avg_time}. We took the runtime of PCA-Keans which is the fastest, as the basic result and give the time cost of the other methods. \textit{Contrario} framework and PCA-Net are implemented with Matlab and tested on CPU. PCA-Kmeans, DSFA and the proposed method are implemented with Python. DSFA and the proposed method are tested on GPU. The CPU used is Intel Core i5-8500 with a clock rate of 3.0 GHz. The GPU used is a single NVIDIA 2080Super card.

\begin{table}[htbp]
    \renewcommand{\arraystretch}{1.3}
    \caption{The comparison of average time cost of different methods}
    \label{tab:avg_time}
    \centering
    \begin{tabular}{c||c|c|c|c|c}
    \hline
     & PCA-Kmeans & DSFA & \textit{Contrario} & PCA-Net & Proposed\\
    \hline\hline
    \bfseries Cost & 1$\times$ & 3.1$\times$ & 12.0$\times$ & 17.6$\times$ & 20.8$\times$\\
    \hline
    \end{tabular}
\end{table}

As observed in Table \ref{tab:avg_time}, PCA-Kmeans is the fastest, then follows the DSFA, which utilized fully connected
network and could be accelerated by GPU. Though the proposed method is the most time consuming, it is still acceptable considering its improvements on unregistered areas. The total runtime of the proposed change detection framework is about 20 minutes. Besides, the total usage of RAM on graphic card is about 1800MB, which makes it possible for training several models in parallel and practicable in realistic scenarios. Once the model training has been finished, the generated images inference and change map generation could be finished in 2 seconds.

\begin{figure}[t!]
    \centering
    \includegraphics[width=3.2in]{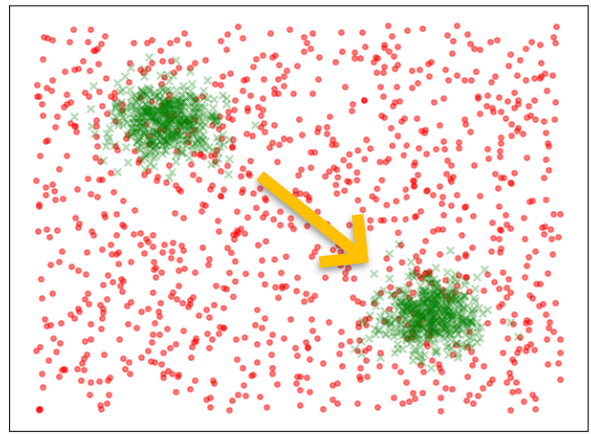}
    \caption{Illustration of the failure at applying proposed method on the data set containing lots of paired satellite images. Red points represent the training data and green points represent the generated images. Since only few red points were used in a mini-batch and objective functions force generated images to be similar to them, it is difficult to make the distribution of generated data approach to the original distribution and the features in green points would be affected by the mini-batches.}
    \label{fig:walk}
\end{figure}

\subsection{Limitations and Future Works}
Although the proposed method can achieve compelling results in many scenes, we could not guarantee it is uniformly positive in all cases. When deep learning architecture is introduced in the algorithm, a model, that based on data sets containing large number of images, is expected.
But unsupervised algorithms always treat every single data as individuals, and GAN looks a little overqualified for handling two images. So we tried our method on a data set containing over 6000 paired images, but after a long time training, model showed no sign of convergence and generator could not produce realistic images. 
This failure might be caused by the diversity of data set. Notice that the term $d(I_0,x_g)+d(I_1,x_g)$ in discriminator's objective function should be lowered, so discriminator would tend to remember the images it saw recently and guide generator to produce pictures with features in images it remembered. The mini-batch that is used for single training iteration could not contain all main features in the image space because the images might describe totally different scenes. As a result, the space of generated images was restricted in a tiny part and kept walking through the whole data space (Fig. \ref{fig:walk}). This problem limited more different images being added into data set.

The final comparison strategy for producing change map might also have a more appropriate one. A common solution is that after getting the optimized generator, fixing its parameters and finding two noise vectors that can be transformed to images which are closest to original two. Then interpolation between two vectors can be applied and all changes would be in control. In this way, the deep learning structure would be fully utilized. But finding two specific points in a continuous space is really difficult and we just leave it along with the discussion in Section \uppercase\expandafter{\romannumeral3}.E for future works.

\section{conclusion}
In this paper, a GAN-based procedure for change detection in satellite images has been proposed. 
The GAN model is designed to produce coregistered images describing the same scenes but differing in changed regions. 
The proposed objective functions enforce Lipschitz constraints that guarantees no gradients vanishing in training process. 
The support expanding strategy creates more training images to handle the problem of sufficient data. 
The structure of discriminator plays a key role in recovering the details in images. 
A simple comparison strategy, which utilizes the outputs of optimized generator, is implemented to obtain the final change map. 
The proposed method demonstrated good performance on generating realistic images and achieved compelling results on the unsupervised change detection task by successfully alleviating the effect of unregistered pixels. 
Another experiments on the structure of discriminator demonstrated the necessity of using clips in suitable size to produce proper images.

There are still several ways our work could be continued, such as how to produce images with better quality and enforce the position information of clips in training stage. And better comparison strategy that utilizes the generated images is also in consideration. Moreover, more change detection tasks, i.e. detecting changes in sequential images or detecting multiple changes, could be explored following the idea of learning suitable representations by learning in the model space \cite{model_space_4, model_space_5} framework.

\ifCLASSOPTIONcaptionsoff
  \newpage
\fi

\bibliographystyle{ieeetr}
\bibliography{ref}

\begin{thebibliography}{10}

\bibitem{cd_1}
R.~J. Radke, S.~Andra, O.~Al{-}Kofahi, and B.~Roysam, ``Image change detection
  algorithms: a systematic survey,'' {\em {IEEE} Trans. Image Process.},
  vol.~14, no.~3, pp.~294--307, 2005.

\bibitem{cd_2}
D.~Lu, P.~Mausel, and E.~F. Moran, ``Change detection techniques,'' {\em
  International Journal of Remote Sensing}, vol.~25, no.~12, pp.~2365--2407,
  2004.

\bibitem{app_1}
P.~Gamba, F.~Dell'Acqua, and G.~Lisini, ``Change detection of multitemporal
  {SAR} data in urban areas combining feature-based and pixel-based
  techniques,'' {\em {IEEE} Trans. Geosci. Remote. Sens.}, vol.~44, no.~10-1,
  pp.~2820--2827, 2006.

\bibitem{app_2}
C.~Marin, F.~Bovolo, and L.~Bruzzone, ``Building change detection in
  multitemporal very high resolution {SAR} images,'' {\em {IEEE} Trans. Geosci.
  Remote. Sens.}, vol.~53, no.~5, pp.~2664--2682, 2015.

\bibitem{app_3}
L.~Giustarini, R.~Hostache, P.~Matgen, G.~J. Schumann, P.~D. Bates, and D.~C.
  Mason, ``A change detection approach to flood mapping in urban areas using
  terrasar-x,'' {\em {IEEE} Trans. Geosci. Remote. Sens.}, vol.~51, no.~4-2,
  pp.~2417--2430, 2013.

\bibitem{app_4}
G.~Moser and S.~B. Serpico, ``Generalized minimum-error thresholding for
  unsupervised change detection from {SAR} amplitude imagery,'' {\em {IEEE}
  Trans. Geosci. Remote. Sens.}, vol.~44, no.~10-2, pp.~2972--2982, 2006.

\bibitem{app_5}
N.~Longbotham, F.~Pacifici, T.~C. Glenn, A.~Zare, M.~Volpi, D.~Tuia,
  E.~Christophe, J.~Michel, J.~Inglada, J.~Chanussot, and Q.~Du, ``Multi-modal
  change detection, application to the detection of flooded areas: Outcome of
  the 2009-2010 data fusion contest,'' {\em {IEEE} J. Sel. Top. Appl. Earth
  Obs. Remote. Sens.}, vol.~5, no.~1, pp.~331--342, 2012.

\bibitem{app_6}
C.~Marin, F.~Bovolo, and L.~Bruzzone, ``Building change detection in
  multitemporal very high resolution {SAR} images,'' {\em {IEEE} Trans. Geosci.
  Remote. Sens.}, vol.~53, no.~5, pp.~2664--2682, 2015.

\bibitem{app_7}
G.~Cao, X.~Li, and L.~Zhou, ``Unsupervised change detection in high spatial
  resolution remote sensing images based on a conditional random field model,''
  {\em Eur. J. Remote Sens.}, vol.~49, no.~1, pp.~225--237, 2016.

\bibitem{camps2008kernel}
G.~Camps-Valls, L.~G{\'o}mez-Chova, J.~Mu{\~n}oz-Mar{\'\i}, J.~L.
  Rojo-{\'A}lvarez, and M.~Mart{\'\i}nez-Ram{\'o}n, ``Kernel-based framework
  for multitemporal and multisource remote sensing data classification and
  change detection,'' {\em {IEEE} Trans. Geosci. Remote. Sens.}, vol.~46,
  no.~6, pp.~1822--1835, 2008.

\bibitem{registration_1}
F.~Pacifici, N.~Longbotham, and W.~J. Emery, ``The importance of physical
  quantities for the analysis of multitemporal and multiangular optical very
  high spatial resolution images,'' {\em {IEEE} Trans. Geosci. Remote. Sens.},
  vol.~52, no.~10, pp.~6241--6256, 2014.

\bibitem{registration_2}
Y.~T.~S. Correa, F.~Bovolo, and L.~Bruzzone, ``An approach for unsupervised
  change detection in multitemporal {VHR} images acquired by different
  multispectral sensors,'' {\em Remote. Sens.}, vol.~10, no.~4, p.~533, 2018.

\bibitem{feature_ext}
B.~Zhou, {\`{A}}.~Lapedriza, J.~Xiao, A.~Torralba, and A.~Oliva, ``Learning
  deep features for scene recognition using places database,'' in {\em Annual
  Conference on Neural Information Processing Systems 2014, Montreal, Quebec,
  Canada, December 8-13}, pp.~487--495, 2014.

\bibitem{learningtransfer}
B.~Zhou, {\`{A}}.~Lapedriza, J.~Xiao, A.~Torralba, and A.~Oliva, ``Learning
  deep features for scene recognition using places database,'' in {\em Annual
  Conference on Neural Information Processing Systems, Montreal, Quebec,
  Canada, December 8-13}, pp.~487--495, 2014.

\bibitem{gan}
I.~Goodfellow, J.~Pouget-Abadie, M.~Mirza, B.~Xu, D.~Warde-Farley, S.~Ozair,
  A.~Courville, and Y.~Bengio, ``Generative adversarial nets,'' in {\em Annual
  Conference on Neural Information Processing Systems, Montreal, Quebec,
  Canada, December 8-13}, pp.~2672--2680, 2014.

\bibitem{unet++}
D.~Peng, Y.~Zhang, and H.~Guan, ``End-to-end change detection for high
  resolution satellite images using improved unet++,'' {\em Remote. Sens.},
  vol.~11, no.~11, p.~1382, 2019.

\bibitem{cgan_cd}
M.~Lebedev, Y.~V. Vizilter, O.~Vygolov, V.~Knyaz, and A.~Y. Rubis, ``Change
  detection in remote sensing images using conditional adversarial networks.,''
  {\em International Archives of the Photogrammetry, Remote Sensing \& Spatial
  Information Sciences}, vol.~42, no.~2, 2018.

\bibitem{cgan}
M.~Mirza and S.~Osindero, ``Conditional generative adversarial nets,'' {\em
  CoRR}, vol.~abs/1411.1784, 2014.

\bibitem{supervised_cd}
Y.~Zhan, K.~Fu, M.~Yan, X.~Sun, H.~Wang, and X.~Qiu, ``Change detection based
  on deep siamese convolutional network for optical aerial images,'' {\em
  {IEEE} Geosci. Remote. Sens. Lett.}, vol.~14, no.~10, pp.~1845--1849, 2017.

\bibitem{cva}
W.~A. Malila, ``Change vector analysis: an approach for detecting forest
  changes with landsat,'' in {\em LARS symposia}, p.~385, 1980.

\bibitem{pca-kmeans}
T.~{\c{C}}elik, ``Unsupervised change detection in satellite images using
  principal component analysis and k -means clustering,'' {\em {IEEE} Geosci.
  Remote. Sens. Lett.}, vol.~6, no.~4, pp.~772--776, 2009.

\bibitem{sfa}
C.~Wu, B.~Du, and L.~Zhang, ``Slow feature analysis for change detection in
  multispectral imagery,'' {\em {IEEE} Trans. Geosci. Remote. Sens.}, vol.~52,
  no.~5, pp.~2858--2874, 2013.

\bibitem{rcva}
F.~Thonfeld, H.~Feilhauer, M.~H. Braun, and G.~Menz, ``Robust change vector
  analysis {(RCVA)} for multi-sensor very high resolution optical satellite
  data,'' {\em Int. J. Appl. Earth Obs. Geoinformation}, vol.~50, pp.~131--140,
  2016.

\bibitem{object-oriented2001}
T.~Blaschke and G.~J. Hay, ``Object-oriented image analysis and scale-space:
  Theory and methods for modeling and evaluating multi-scale landscape
  structure,'' {\em ISPRS - International Archives of the Photogrammetry,
  Remote Sensing and Spatial Information Sciences}, pp.~22--29, 2001.

\bibitem{object-based2008}
J.~Im, J.~Jensen, and J.~Tullis, ``Object-based change detection using
  correlation image analysis and image segmentation,'' {\em International
  Journal of Remote Sensing}, vol.~29, no.~2, pp.~399--423, 2008.

\bibitem{object-based2012}
G.~Chen, G.~J. Hay, L.~M.~T. De~Carvalho, and M.~A. Wulder, ``Object-based
  change detection,'' {\em International Journal of Remote Sensing}, vol.~33,
  no.~14, pp.~4434--4457, 2012.

\bibitem{object-based2018}
X.~Wang, S.~Liu, P.~Du, H.~Liang, J.~Xia, and Y.~Li, ``Object-based change
  detection in urban areas from high spatial resolution images based on
  multiple features and ensemble learning,'' {\em Remote. Sens.}, vol.~10,
  no.~2, p.~276, 2018.

\bibitem{ocva}
L.~Li, X.~Li, Y.~Zhang, L.~Wang, and G.~Ying, ``Change detection for
  high-resolution remote sensing imagery using object-oriented change vector
  analysis method,'' in {\em {IEEE} International Geoscience and Remote Sensing
  Symposium, Beijing, China, July 10-15}, pp.~2873--2876, 2016.

\bibitem{contrario}
G.~Liu, Y.~Gousseau, and F.~Tupin, ``A contrario comparison of local
  descriptors for change detection in very high spatial resolution satellite
  images of urban areas,'' {\em {IEEE} Trans. Geosci. Remote. Sens.}, vol.~57,
  no.~6, pp.~3904--3918, 2019.

\bibitem{sift}
P.~C. Ng and S.~Henikoff, ``Sift: predicting amino acid changes that affect
  protein function,'' {\em Nucleic Acids Research}, vol.~31, no.~13,
  pp.~3812--3814, 2003.

\bibitem{bipartite}
J.~Liu, M.~Gong, A.~K. Qin, and K.~C. Tan, ``Bipartite differential neural
  network for unsupervised image change detection,'' {\em {IEEE} Trans. Neural
  Networks Learn. Syst.}, vol.~31, no.~3, pp.~876--890, 2020.

\bibitem{dsfa}
B.~Du, L.~Ru, C.~Wu, and L.~Zhang, ``Unsupervised deep slow feature analysis
  for change detection in multi-temporal remote sensing images,'' {\em {IEEE}
  Trans. Geosci. Remote. Sens.}, vol.~57, no.~12, pp.~9976--9992, 2019.

\bibitem{unsupervisedCVA}
S.~Saha, F.~Bovolo, and L.~Bruzzone, ``Unsupervised deep change vector analysis
  for multiple-change detection in vhr images,'' {\em {IEEE} Trans. Geosci.
  Remote. Sens.}, vol.~57, no.~6, pp.~3677--3693, 2019.

\bibitem{autoencoders}
N.~Lv, C.~Chen, T.~Qiu, and A.~K. Sangaiah, ``Deep learning and superpixel
  feature extraction based on contractive autoencoder for change detection in
  {SAR} images,'' {\em {IEEE} Trans. Ind. Informatics}, vol.~14, no.~12,
  pp.~5530--5538, 2018.

\bibitem{model_space_3}
H.~Chen, F.~Tang, P.~Ti{\~{n}}o, A.~G. Cohn, and X.~Yao, ``Model metric
  co-learning for time series classification,'' in {\em Proceedings of the
  Twenty-Fourth International Joint Conference on Artificial Intelligence,
  Buenos Aires, Argentina, July 25-31}, pp.~3387--3394, 2015.

\bibitem{model_space_1}
H.~Chen, P.~Ti{\~{n}}o, A.~Rodan, and X.~Yao, ``Learning in the model space for
  cognitive fault diagnosis,'' {\em {IEEE} Trans. Neural Networks Learn.
  Syst.}, vol.~25, no.~1, pp.~124--136, 2014.

\bibitem{model_space_2}
H.~Chen, F.~Tang, P.~Ti{\~{n}}o, and X.~Yao, ``Model-based kernel for efficient
  time series analysis,'' in {\em The 19th {ACM} International Conference on
  Knowledge Discovery and Data Mining, Chicago, IL, USA, August 11-14},
  pp.~392--400, 2013.

\bibitem{f-gan}
S.~Nowozin, B.~Cseke, and R.~Tomioka, ``f-gan: Training generative neural
  samplers using variational divergence minimization,'' in {\em Annual
  Conference on Neural Information Processing Systems, Barcelona, Spain,
  December 5-10}, pp.~271--279, 2016.

\bibitem{wgan-former}
M.~Arjovsky and L.~Bottou, ``Towards principled methods for training generative
  adversarial networks,'' in {\em 5th International Conference on Learning
  Representations, Toulon, France, April 24-26}, 2017.

\bibitem{wgan}
M.~Arjovsky, S.~Chintala, and L.~Bottou, ``Wasserstein generative adversarial
  networks,'' in {\em Proceedings of the 34th International Conference on
  Machine Learning, Sydney, NSW, Australia, 6-11 August}, pp.~214--223, 2017.

\bibitem{wgan-gp}
I.~Gulrajani, F.~Ahmed, M.~Arjovsky, V.~Dumoulin, and A.~C. Courville,
  ``Improved training of wasserstein gans,'' in {\em Annual Conference on
  Neural Information Processing Systems, Long Beach, CA, {USA}, 4-9 December},
  pp.~5767--5777, 2017.

\bibitem{spectralnorm}
T.~Miyato, T.~Kataoka, M.~Koyama, and Y.~Yoshida, ``Spectral normalization for
  generative adversarial networks,'' in {\em 6th International Conference on
  Learning Representations, Vancouver, BC, Canada, April 30 - May 3}, 2018.

\bibitem{gan-div}
J.~Wu, Z.~Huang, J.~Thoma, D.~Acharya, and L.~Van~Gool, ``Wasserstein
  divergence for gans,'' in {\em 15th European Conference, Munich, Germany,
  September 8-14}, pp.~673--688, 2018.

\bibitem{ganqp}
J.~Su, ``{GAN-QP:} {A} novel {GAN} framework without gradient vanishing and
  lipschitz constraint,'' {\em CoRR}, vol.~abs/1811.07296, 2018.

\bibitem{shift-variance}
R.~Zhang, ``Making convolutional networks shift-invariant again,'' in {\em
  Proceedings of the 36th International Conference on Machine Learning, Long
  Beach, California, USA, June 9-15} (K.~Chaudhuri and R.~Salakhutdinov, eds.),
  pp.~7324--7334, 2019.

\bibitem{pix2pix}
P.~Isola, J.~Zhu, T.~Zhou, and A.~A. Efros, ``Image-to-image translation with
  conditional adversarial networks,'' in {\em {IEEE} Conference on Computer
  Vision and Pattern Recognition, Honolulu, HI, USA, July 21-26},
  pp.~5967--5976, 2017.

\bibitem{2018patch}
U.~Demir and G.~B. {\"{U}}nal, ``Patch-based image inpainting with generative
  adversarial networks,'' {\em CoRR}, vol.~abs/1803.07422, 2018.

\bibitem{mar_gan}
C.~Li and M.~Wand, ``Precomputed real-time texture synthesis with markovian
  generative adversarial networks,'' in {\em 14th European Conference,
  Amsterdam, The Netherlands, October 11-14}, pp.~702--716, 2016.

\bibitem{mar_gan2}
C.~Li and M.~Wand, ``Combining markov random fields and convolutional neural
  networks for image synthesis,'' in {\em {IEEE} Conference on Computer Vision
  and Pattern Recognition, Las Vegas, NV, USA, June 27-30}, pp.~2479--2486,
  2016.

\bibitem{mar_gan3}
L.~A. Gatys, A.~S. Ecker, and M.~Bethge, ``Texture synthesis and the controlled
  generation of natural stimuli using convolutional neural networks,'' in {\em
  Bernstein Conference}, pp.~219--219, 2015.

\bibitem{Adam}
D.~P. Kingma and J.~Ba, ``Adam: {A} method for stochastic optimization,'' in
  {\em 3rd International Conference on Learning Representations, San Diego, CA,
  USA, May 7-9} (Y.~Bengio and Y.~LeCun, eds.), 2015.

\bibitem{batch_norm}
S.~Ioffe and C.~Szegedy, ``Batch normalization: Accelerating deep network
  training by reducing internal covariate shift,'' in {\em Proceedings of the
  32nd International Conference on Machine Learning, Lille, France, July 6-11}
  (F.~R. Bach and D.~M. Blei, eds.), pp.~448--456, 2015.

\bibitem{pca-net}
F.~Gao, J.~Dong, B.~Li, and Q.~Xu, ``Automatic change detection in synthetic
  aperture radar images based on pcanet,'' {\em {IEEE} Geosci. Remote. Sens.
  Lett.}, vol.~13, no.~12, pp.~1792--1796, 2016.

\bibitem{FID}
M.~Heusel, H.~Ramsauer, T.~Unterthiner, B.~Nessler, and S.~Hochreiter, ``Gans
  trained by a two time-scale update rule converge to a local nash
  equilibrium,'' in {\em Annual Conference on Neural Information Processing
  Systems, Long Beach, CA, {USA}, December 4-9}, pp.~6626--6637, 2017.

\bibitem{aux-learn}
A.~Valada, N.~Radwan, and W.~Burgard, ``Deep auxiliary learning for visual
  localization and odometry,'' in {\em {IEEE} International Conference on
  Robotics and Automation, Brisbane, Australia, May 21-25}, pp.~6939--6946,
  2018.

\bibitem{model_space_4}
Z.~Gong and H.~Chen, ``Model-based oversampling for imbalanced sequence
  classification,'' in {\em Proceedings of the 25th {ACM} International
  Conference on Information and Knowledge Management, Indianapolis, IN, USA,
  October 24-28}, pp.~1009--1018, 2016.

\bibitem{model_space_5}
Z.~Gong, H.~Chen, B.~Yuan, and X.~Yao, ``Multiobjective learning in the model
  space for time series classification,'' {\em {IEEE} Trans. Cybern.}, vol.~49,
  no.~3, pp.~918--932, 2019.

\end{thebibliography}

\end{document}